# Hybrid aeromaterials for enhanced and rapid volumetric photothermal response

Lena M. Saure[1], Niklas Kohlmann[2], Haoyi Qiu[1], Shwetha Shetty[3], Ali Shaygan Nia[4], Narayanan Ravishankar[3], Xinliang Feng[4], Alexander Szameit[5], Lorenz Kienle[2,6], Rainer Adelung[1,6], Fabian Schütt[1,6,*]

[1] Functional Nanomaterials, Department for Materials Science, Kiel University, Kaiser Str. 2, 24143 Kiel, Germany

[2] Synthesis and Real Structure, Department for Materials Science, Kiel University, Kaiser Str. 2, 24143 Kiel, Germany

[3] Materials Research Centre, Indian Institute of Science, Bangalore, India

[4] Department of Chemistry and Food Chemistry, Center for Advancing Electronics Dresden (cfaed), Technische Universität Dresden, 01062 Dresden, Germany

[5] Institut für Physik, Universität Rostock, 18059 Rostock, Germany

[6] Kiel Nano, Surface and Interface Science KiNSIS, Kiel University, Christian-Albrechts-Platz 4, 24118 Kiel, Germany

[*] Corresponding author: Fabian Schütt, fas@tf.uni-kiel.de

**Abstract**

Conversion of light into heat is essential for a broad range of technologies such as solar thermal heating, catalysis and desalination. Three-dimensional (3D) carbon nanomaterial-based aerogels have shown to hold great promise as photothermal transducer materials. However, till now, their light-to-heat conversion is limited by surface-near absorption, resulting in a strong heat localization only at the illuminated surface region, while most of the aerogel volume remains unused. We present an innovative fabrication concept for highly porous (>99.9%) photothermal hybrid aeromaterials, that enable an ultra-rapid and volumetric photothermal response with an enhancement by a factor of around 2.5 compared to the pristine variant. The hybrid aeromaterial is based on strongly light-scattering framework structures composed of interconnected hollow silicon dioxide ($SiO_2$) microtubes, which are functionalized with extremely low amounts (in order of a few µg $cm^{-3}$) of reduced graphene oxide (rGO) nanosheets, acting as photothermal agents. Tailoring the density of rGO within the framework structure enables us to control both, light scattering and light absorption, and thus the volumetric photothermal response. We further show that by rapid and repeatable gas activation these transducer materials expand the field of photothermal applications, like untethered light-powered and -controlled microfluidic pumps and soft pneumatic actuators.

**Table of contents:**

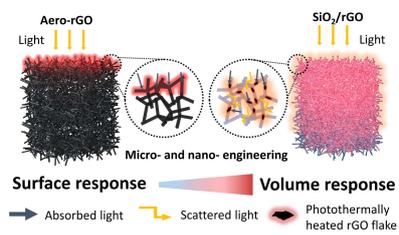

Innovative hybrid aeromaterials allow to overcome the fundamental limits of solid surface absorbers and show a rapid and enhanced volumetric photothermal response. Repeatable activation of macroscopic gas volumes by volumetric light-to-heat conversion enables applications like untethered light-powered and -controlled microfluidic pumps and soft pneumatic actuators.

# 1. Introduction

The generation of thermal energy from light is an essential physical process that does not only govern all biological processes, but is of major importance for a broad variety of technologies, ranging from fundamental environmental applications, like hot water generation[1], desalination[1], thermal management of buildings[2], power generation[1], solar thermal heating[3], catalysis[4] to medical applications[5]. For the conversion of light to heat, the photothermal effect is exploited, where light is absorbed by photothermal materials, resulting in a temperature rise. The photothermal efficiency, i.e. the amount of light converted into thermal energy, strongly depends on the type of material and its surface properties. Only light that is not directly reflected can be either converted into electric energy (photovoltaic effect), into another wavelength (fluorescence) or into thermal energy. The latter usually accounts for the largest part.

With respect to that, nanomaterials have been shown to hold great potential as efficient photothermal agents, due to their large surface-to-volume ratio, as well as their unique electronic structure. Photothermal nanomaterials include semiconductor nanomaterials, plasmonic metal nanoparticles, polymer-based nanomaterials, as well as carbon and related nanomaterials, such as Graphene, Carbon Nanotubes and Mxenes.[6–9] For example, just recently, a photothermal efficiency of ~100% could be reached by utilizing $Ti_3C_2$ based nanomaterials.[10] Due to their small size (below 100 nm), photothermal nanomaterials have been employed as therapeutic agents for the localized heat treatment of cancer[5], as well as in the sterilization of surfaces to prevent bacterial biofilm formation[11].

Moreover, three-dimensional (3D) aerogels and foam structures made of the aforementioned photothermal nanomaterials, have recently been shown to hold great potential as light-to-heat transducer materials. Their extremely low volumetric density (typically below 20 mg cm$^{-3}$)[12,13], results in a volumetric heat capacity similar to that of air, enabling a rapid and efficient heating under light illumination. Simultaneously, their macroscopic size, their porous structure (typically > 98%)[12,13] and high surface area (up to 75 m² g$^{-1}$)[13] lead to a unique set of light–matter interactions, including light scattering, light absorption and reflection. Their potential as an excellent material platform for light-to-heat conversion, has already been demonstrated for several applications, ranging from photoacoustics[14], water steam generation[7–9,13], hydrogen catalysis[15] as well as desalination[1,7].

However, while these studies indicate the potential of carbon and related nanomaterial-based foams as light-to-heat transducer materials, in all reported cases, the light illumination only results in a strong heat localization at the surface region of the photothermal active aerogel[8,9,13], as the light penetration into the macroscopic aerogel structure is physically limited. Especially, the transfer of heat to the surrounding medium, e.g. gas, which is essential for most applications, is physically limited by the low density of gases (typically 4 orders of magnitude below solids), requiring a high surface area for activation of large gas volumes.

Here we present a new concept for the design and development of nano- and micro-engineered hybrid aeromaterial transducers with enhanced volumetric light-to-heat conversion. The concept is based on an open porous (porosity ~ 99.9%), ultralightweight (density ~ 3 mg cm$^{-3}$), non-absorbing and light scattering framework structure composed of silicon dioxide ($SiO_2$) microtubes, named aeroglass (AG), functionalized with low volumetric loadings (between 4.7 µg cm$^{-3}$ and 94.7 µg cm$^{-3}$) of reduced graphene oxide (rGO), acting as the photothermal agent. By utilizing the light scattering properties of the pristine AG, and tailoring the volumetric density of rGO within the AG, we show, that the light penetration depth into the foams can be increased up to 10 millimeters, thereby increasing the photothermally activated gas volume by a factor up to 2.44 compared to a pristine rGO aeromaterial transducer, which, due to a high open porosity, already has a relatively high light penetration depth compared to conventional aerogels. We further demonstrate, that these micro- and nano-engineered transducer materials enable completely new application scenarios for light-to-heat conversion, including untethered and light-driven pneumatic actuators for soft robotics, as well as wireless light-powered and -controlled microfluidic pumps.

## 2. Results
### 2.1 Synthesis and characterization

The synthesis process of the macroscopic nano- and micro-engineered transducer material is schematically shown in **Figure 1a**. In short, a sacrificial network of tetrapodal-shaped ZnO microparticles is first partially coated with photothermal nanomaterial by a simple infiltration process, followed by a wet chemical $SiO_2$ coating step. Etching of ZnO and supercritical drying results in a freestanding multi-scale material system composed of interconnected hollow $SiO_2$ microtubes decorated with photothermal nanomaterial. More details on the synthesis can be found in the materials and methods section. We here use reduced graphene oxide (rGO) as a light absorbing nanomaterial due to the broadband absorption of UV-vis and near infrared range (NIR) light.[5,6,13,16] However, other light absorbing nanomaterials such as gold nanoparticles can be incorporated by the same method as well, see **Figure S1**. By adjusting the concentration of the used graphene oxide (GO) dispersion, the volumetric loading of rGO within the AG can be precisely adjusted between 0 and 94.7 µg cm$^{-3}$, which is key to tailor the volumetric light–matter interaction of the system. **Figure 1b** shows a photograph of cylindrical AG/rGO samples with different volumetric loadings of rGO, as well as pristine AG and the pristine rGO variant, named Aero-rGO. While pristine AG shows a white appearance, a color shift to dark grey with increasing rGO loading in the AG/rGO hybrid aeromaterial structure is observed. The optical appearance of the pristine Aero-rGO is significantly darker, as the concentration of rGO is 100 times higher (9465 µg cm$^{-3}$) than for the AG/rGO with highest loading of rGO (94.7 µg cm$^{-3}$). Detailed characterization of Aero-rGO can be found in previous publications[17,18]. It has to be noted that the properties, e.g. light transport, of the aeromaterials strongly differ from those of conventional aerogels.[18,19] SEM characterization (**Figure 1c, d**) reveals the open porous network structure as well as the hollow character of the SiO$_2$ microtubes and clearly shows the homogenous distribution of rGO flakes on the surface of the microtubes. A more detailed SEM characterization can be found in **Figure S2**. Individual microtubes have a mean length of ~25 µm and a diameter of ~1-3 µm. The wall thickness of the hollow microtubes was determined by TEM with an average wall thickness of ~17 nm (for more details see Supplementary Note 1). **Figure 1e** shows a SiO$_2$ microtube arm decorated with a rGO flake. EDX mapping confirms the existence of rGO flakes incorporated into the SiO$_2$ layer (**Figure 1f**). The as synthesized SiO$_2$ is amorphous, as determined by TEM investigations (see **Figure S3**). This agrees with the Raman measurements shown in **Figure 1g**, which were conducted for the pristine AG, pristine Aero-rGO and for the highest (AG/rGO-94.7) and lowest (AG/rGO-4.7) degree of rGO functionalization. Typical Raman peaks for rGO are observed in both the AG/rGO hybrid material systems as well as the pristine Aero-rGO. The positions of D band (1339 cm$^{-1}$) and G band (1597 cm$^{-1}$) are in accordance with the literature, as well as the less pronounced 2D peak.[17,20] The Raman spectrum of the pristine AG shows distinct peaks at 490 cm$^{-1}$, 800 cm$^{-1}$ and 975 cm$^{-1}$ which are typical for amorphous SiO$_2$ structures.[21]

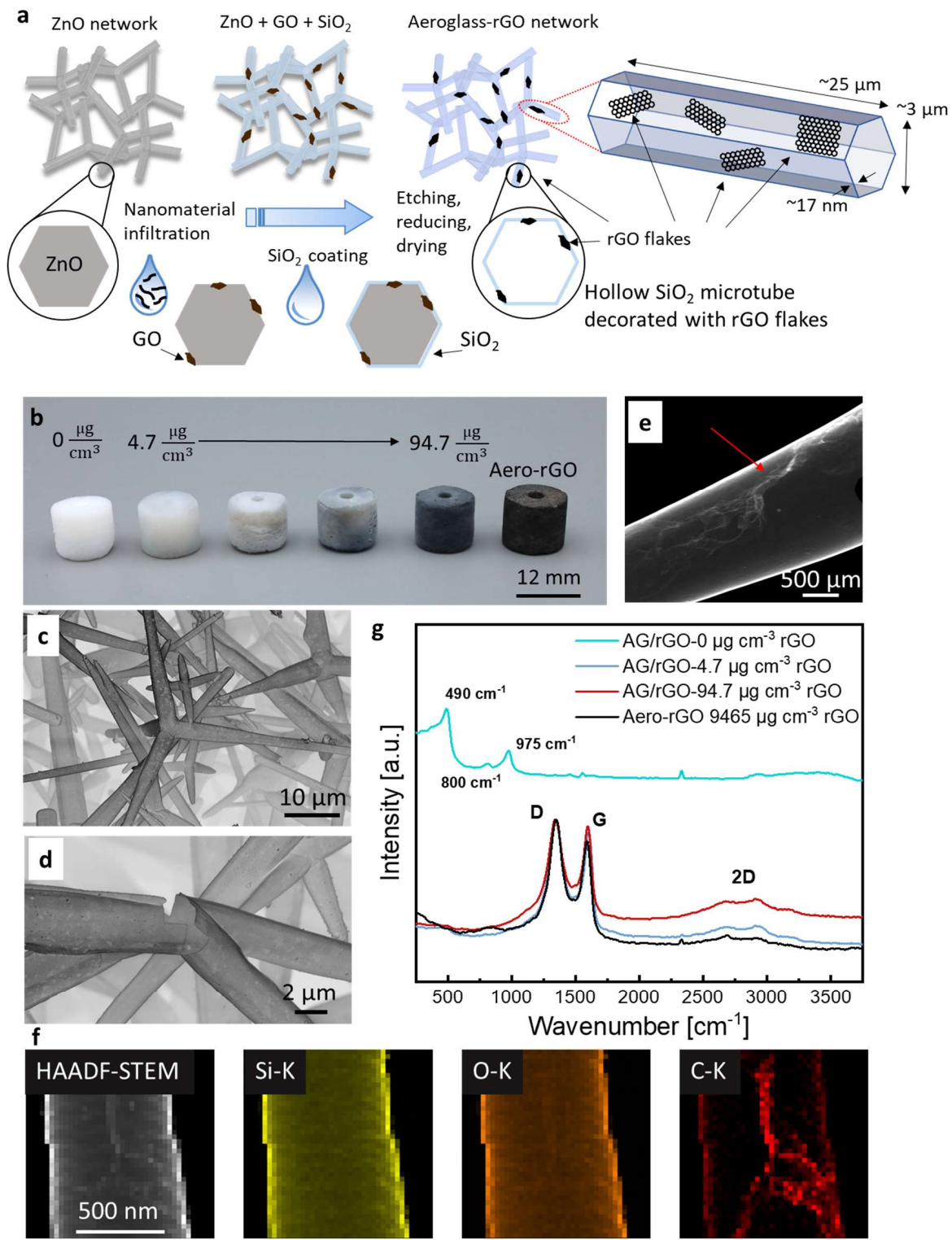

**Figure 1.** Synthesis and characterization of AG/rGO transducer materials. (a) Schematic of synthesis steps of aeroglass functionlized with rGO and magnified hollow microtube with rGO flakes distributed on the surface of the microtube. (b) Photograph of functionalized aeroglass (AG/rGO) samples with different volumetric loadings of rGO (c),(d) SEM images of a AG/rGO-94.7 sample, revealing the open porous structure and the hollow character of microtubes with a homogeneous distribution of rGO flakes. (e) TEM image of a AG/rGO-94.7

sample showing a rGO flake on a SiO$_2$ microtube. (f) EDX mapping of AG/rGO-94.7 for silicon, oxygen and carbon, clearly indicating the existence of a rGO flake on a microtube. (g) Raman spectra for pristine AG, AG/rGO-4.7 and AG/rGO-94.7, as well as for Aero-rGO. The spectra clearly indicate typical peaks for D, G and 2D band for rGO.

## 2.2 Optimizing volumetric light absorption

The as described multi-scale nature of the here used hybrid aeromaterials enables a unique set of light–matter interactions, as shown in **Figure 2a,b**. Illuminating aeromaterial structures of pristine AG (density ~3 mg cm$^{-3}$) with a focused light beam results in strong isotropic light scattering in all room directions, similar to what has been previously reported for framework structures based on hexagonal boron nitride.[19] The AG acts as a volumetric light diffuser. In contrast to that, aeromaterials composed of rGO with a density of 11 mg cm$^{-3}$ absorb all light when illuminated with a focused light beam due to their broadband absorption properties (**Figure 2b**). While these demonstrate the two extreme cases, either complete light scattering in case of pristine AG or complete light absorption in case of pristine Aero-rGO, tailoring the volumetric density of rGO inside of the AG framework structure enables us to adjust the volumetric light–matter interactions of the AG/rGO hybrid aeromaterials in a controlled manner. **Figure 2c** shows the light transmission through cylindrically shaped (d=12 mm) AG/rGO hybrid aeromaterials with varying degree of volumetric rGO density (0 µg cm$^{-3}$, 4.7 µg cm$^{-3}$ and 94.7 µg cm$^{-3}$) as well as Aero-rGO as a function of sample height (2 mm – 10 mm). In the case of pristine Aero-rGO, even at a sample height of 2 mm no light is transmitted through the aeromaterial structure. AG/rGO-94.7 transmits a small amount of light for sample dimension of 2 mm and 4 mm (3.5 % and 1.8 %), whereas the highest transmitted light intensity is detected for pristine AG with 2 mm height (17 %). However, also AG/rGO-4.7 shows high light transmission for 2 mm (15.7 %) which strongly decreases with increasing sample thickness. Even for 10 mm sample thickness a small amount of transmitted light is still detected (2.4 %), meaning that some light can transmit through the structure without being absorbed by rGO. The results indicate that the volumetric density of rGO within the hybrid aeromaterial strongly influences the light penetration depth into the structure, and by this, the size of the light–matter interaction volume. This can also be visualized by thermography, as light interacting with the rGO inside the AG/rGO hybrid aeromaterial results in heat generation by means of the photothermal effect. **Figure 2d** displays the maximum front- and backside temperature extracted from infrared (IR) images for different AG/rGO hybrid aeromaterials (cylindrical geometry, h= 10 mm and d= 12 mm) taken from the frontside (side that is illuminated with light) and backside (see **Figure S4** for IR images). All aeromaterials were illuminated with a broadband light source (details see materials and methods) for only 1 s with an irradiance of 1.67 W cm$^{-2}$, in order to reduce the effect of heat conduction and accumulation. The short illumination time also demonstrates the rapid photothermal heating effect. Due to the negligible absorption properties of pristine AG, see UV-vis spectroscopy results in **Figure 2e**, no increase in temperature in the entire volume of the AG can be detected under illumination. In contrast to that, the pristine Aero-rGO, heats up only at the illuminated front side, meaning that all light is directly absorbed in the surface region of the aeromaterial structure, confirming the light transmission measurements shown in **Figure 2c**. As can be seen in **Figure 2d**, the maximum frontside temperature increases with increasing volumetric rGO loading in the AG/rGO hybrid aeromaterials. UV-vis measurements confirm, that the higher the volumetric rGO density, the higher the absorption over the whole wavelength spectrum (see **Figure 2e**, more details in SI Note 2), resulting in a higher frontside temperature as detected by thermography. In contrast, the backside temperature, as a measure for light penetration depth, reaches a maximum value of 66°C for AG/rGO-47.3. For higher concentrations of rGO (94.7 µg cm$^{-3}$) the backside temperature decreases to 55 °C, whereas for pristine Aero-rGO no significant increase in temperature at the backside of the sample can be observed. This is highlighted as well in IR images (see **Figure 2f**) of the sample backside of AG/rGO-4.7, -94.7 and Aero-rGO with varying sample height (4 mm – 10 mm). The maximum temperature is extracted and displayed in **Figure 2g**. For the hybrid aeromaterial transducers (AG/rGO-94.7 and -4.7) a heating effect in the entire volume can be detected, while for Aero-rGO the volumetric heating effect is already drastically reduced for 4 mm sample thickness. However, for AG/rGO-4.7 the volumetric heating effect is less pronounced, as the low rGO loading results in less absorption.

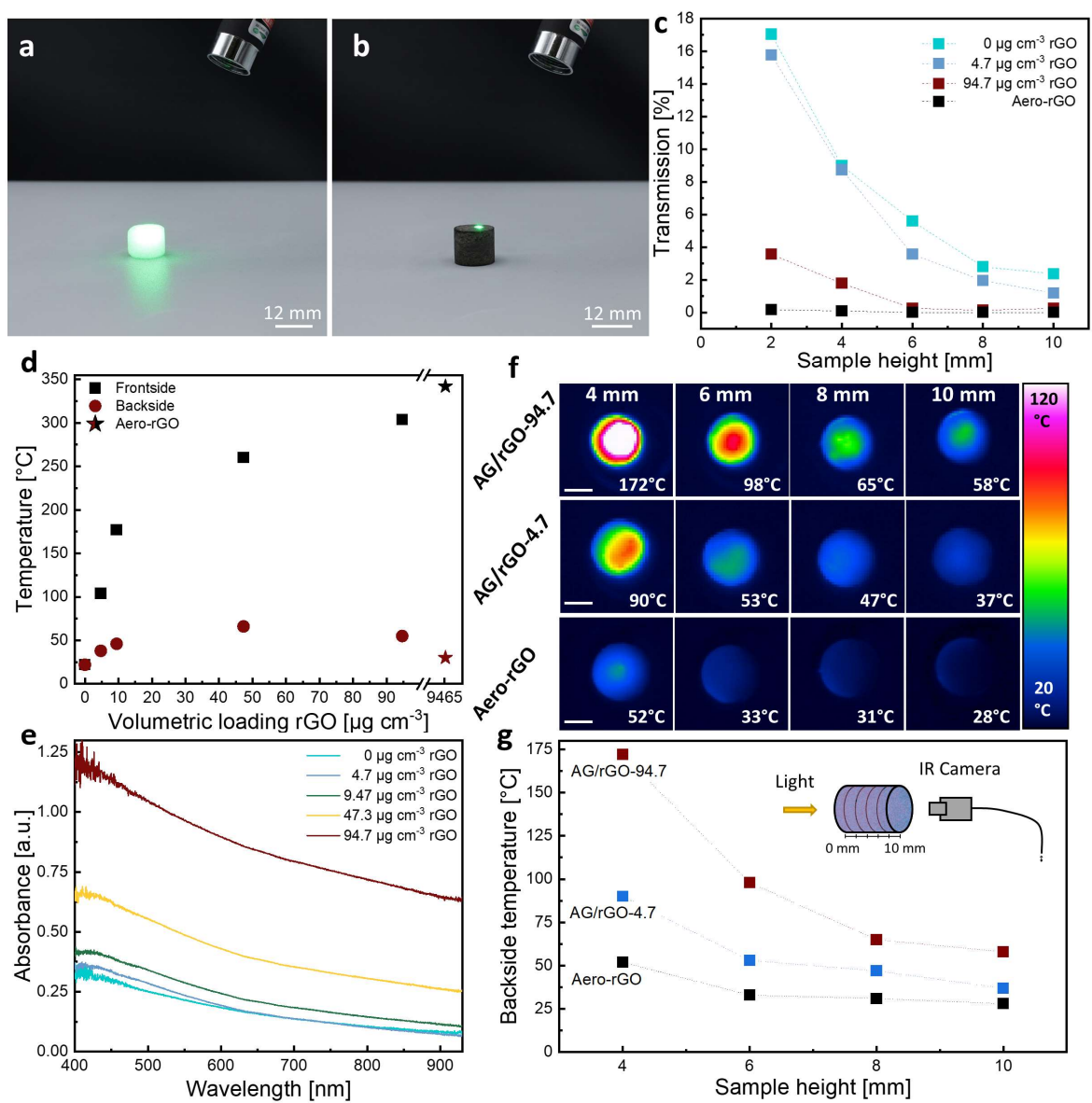

**Figure 2.** Light induced heating of AG/rGO. (a,b) Photograph of pristine AG and pristine Aero-rGO illuminated with a green laser pointer. (c) Light transmission of pristine AG, Aero-rGO and AG/rGO with high (94.7 µg cm$^{-3}$) and low (4.7 µg cm$^{-3}$) volumetric rGO loading measured for different sample heights. (d) Maximum temperature of AG/rGO samples with different volumetric loadings of rGO and a height of 10 mm, illuminated for 1 s with an irradiance of 1.67 W cm$^{-2}$ extracted from IR images of frontside (illumination side) and backside. (e) UV-vis absorbance for different volumetric loadings of rGO within AG/rGO hybrid aeromaterial. (f) IR images of sample backside of AG/rGO with different volumetric rGO loadings and Aero-rGO samples with increasing sample heights, showing the temperature profile after illuminating the samples for 1 s with an irradiance of 1.67 W cm$^{-2}$. Scale bar is 6 mm. (g) Maximum backside temperature for different sample heights. The schematic illustrates the measurement set-up.

These results clearly demonstrate, that an optimized volumetric rGO density in combination with a light-scattering material system can adjust the interaction volume of light within the AG/rGO hybrid aeromaterials, enabling a strong enhancement in volumetric light-to-heat conversion. The proposed concept of volumetric light-to-heat conversion is shown schematically in **Figure 3**, demonstrating the light–matter interaction for three different cases, namely pristine AG, Aero-rGO and AG/rGO with optimized volumetric loading of rGO. Light impinging on pristine AG is scattered multiple times within the structure, before leaving it again with only a negligible amount of light being absorbed, resulting in an isotropic light distribution, as shown in **Figure 2a**. The reason for multiple scattering events taking place can be found in the unique hierarchical microstructure, creating an optical disorder system with many Rayleigh scattering centers, as explained in more detail in Supporting Note 3. Thus, pristine AG is acting solely as a passive light diffuser and no light is converted into heat. In contrast, Aero-rGO (**Figure 3b**) absorbs all impinging light directly at the illumination side resulting in strong localized heating of the microtubular structure. In contrast, AG/rGO with optimized volumetric loading of rGO (**Figure 3c**) enables a combination of light scattering properties with the photothermal effect of the rGO flakes. Due to the homogenous distribution of rGO flakes in the AG framework structure, only a small amount of light is directly absorbed at the illumination side and most of the impinging light can penetrate deep into the structure by multiple scattering events before eventually impinging on a rGO flake, where light is absorbed and converted into heat by the photothermal effect. However, for too low volumetric loadings of rGO (e.g. 4.7 µg cm$^{-3}$) a noticeable amount of light can escape the aeromaterial structure without interacting with the photothermal nanomaterial. Therefore, an optimized volumetric density of rGO inside the AG is necessary to balance between light scattering, transmission and absorbance.

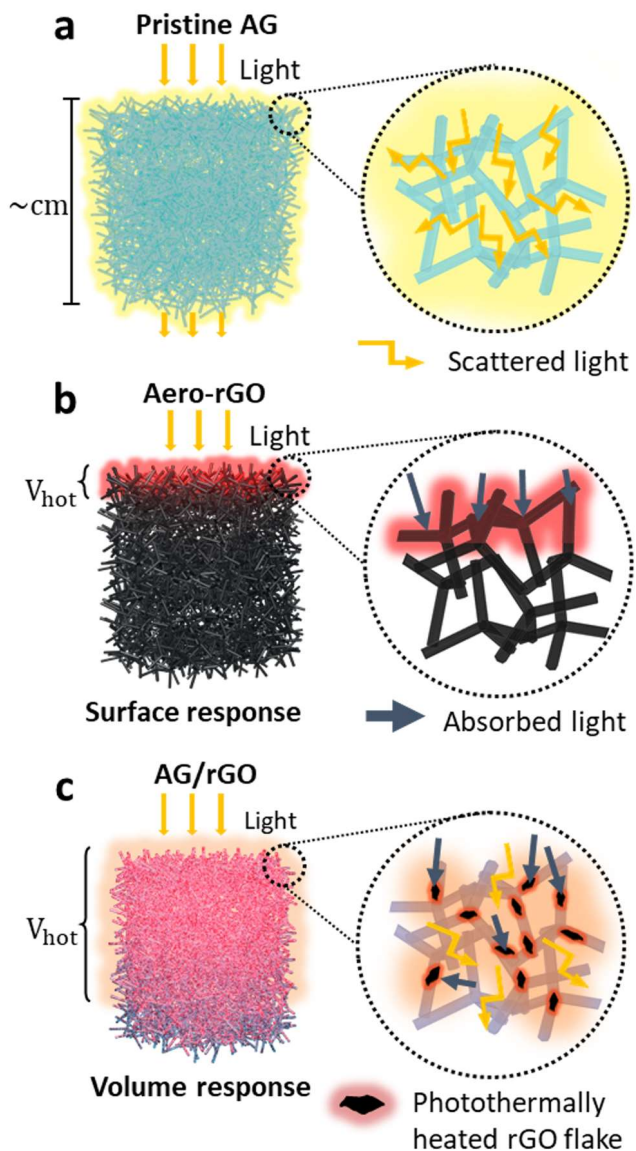

**Figure 3.** Schematic concept of proposed light–matter interaction in aeromaterials. Light scattering, associated light penetration, and light absorption resulting in diffferent photothermally activce volumes for (a) pristine AG, (b) pristine Aero-rGO and (c) AG/rGO.

## 2.3 Activation of gas volume

The enhanced volumetric light absorption in combination with the extremely low volumetric heat capacity (~ 2.5 kJ m$^{-3}$ K$^{-1}$) (Calculation details in SI Note 4) and high gravimetric surface area (~ 74.153 m$^2$ g$^{-1}$) (Calculation details in SI Note 5) of the here developed nano- and micro-engineered transducer material enables a rapid and repeatable activation (heating) of macroscopic gas volumes, as schematically shown in **Figure 4a**. Based on the low volumetric heat capacity, the temperature of the AG/rGO hybrid aeromaterial is rapidly increasing when illuminated with light, i.e. in only 1 s the temperature increases to around 300°C (for AG/rGO-

94.7). The heat generated by the photothermal effect within the aeromaterial structure is transferred to the gas phase within milliseconds by utilizing its high surface area. Under isobaric conditions, this causes a rapid volume expansion of the gas within the aeromaterial, according to ideal gas law. Simultaneously, the open porous framework structure (porosity >99.9 %) with micrometer-sized voids, allows rapid gas exchange out of and into the aeromaterial structure. Without illumination, the structure cools down immediately, and the process can be repeated. A similar effect has been reported before, based on the Joule-heating of framework structures of graphene [18], however, electrical contacts are required, whereas here, we present wireless activation of macroscopic gas volumes using light.

In order to further quantify the volumetric light-to-heat conversion and related rapid gas activation of our hybrid aeromaterial transducer as a function of volumetric rGO loading, illumination time and irradiance, AG/rGO hybrid aeromaterials were placed in an air-tight chamber equipped with two check valves, as shown in the inset of **Figure 4b**, and in **Figure S5** and **Figure S6** In this configuration the illumination of an AG/rGO hybrid aeromaterial through a polymethylmethacrylate window results in a rapid temperature increase of the gas within the sample volume, causing a volume expansion. The extent of the volume expansion, according to the ideal gas law, thereby depends on the temperature and the amount of activated gas and thus on the properties of the transducer material. The check valves allow directional gas flow through a connected glass tube containing 1 ml of dyed water. By applying a light pulse, the water droplet is pushed through the glass tube (see Supplementary Video 1), with the distance being a direct measure of the activated gas volume.

The mean activated gas volume for one light pulse of 1 s for different volumetric rGO loadings is displayed in **Figure 4b** (irradiance of 1.24 W cm$^{-2}$, f = 0.5 Hz, more details see materials and methods). In accordance to the results presented in **Figure 2**, there is no gas activated in the case of pristine AG, as only negligible light absorption occurs. With increasing volumetric loading of rGO, the activated gas volume increases to a maximum value of 0.11 cm$^3$ for AG/rGO-47.3, which is around 9.7 % of the total volume of the aeromaterial structure. Compared to pristine Aero-rGO this means an improvement by a factor of 2.44 in terms of volumetric gas activation, as for Aero-rGO only a gas volume of around 0.045 cm³ is activated.

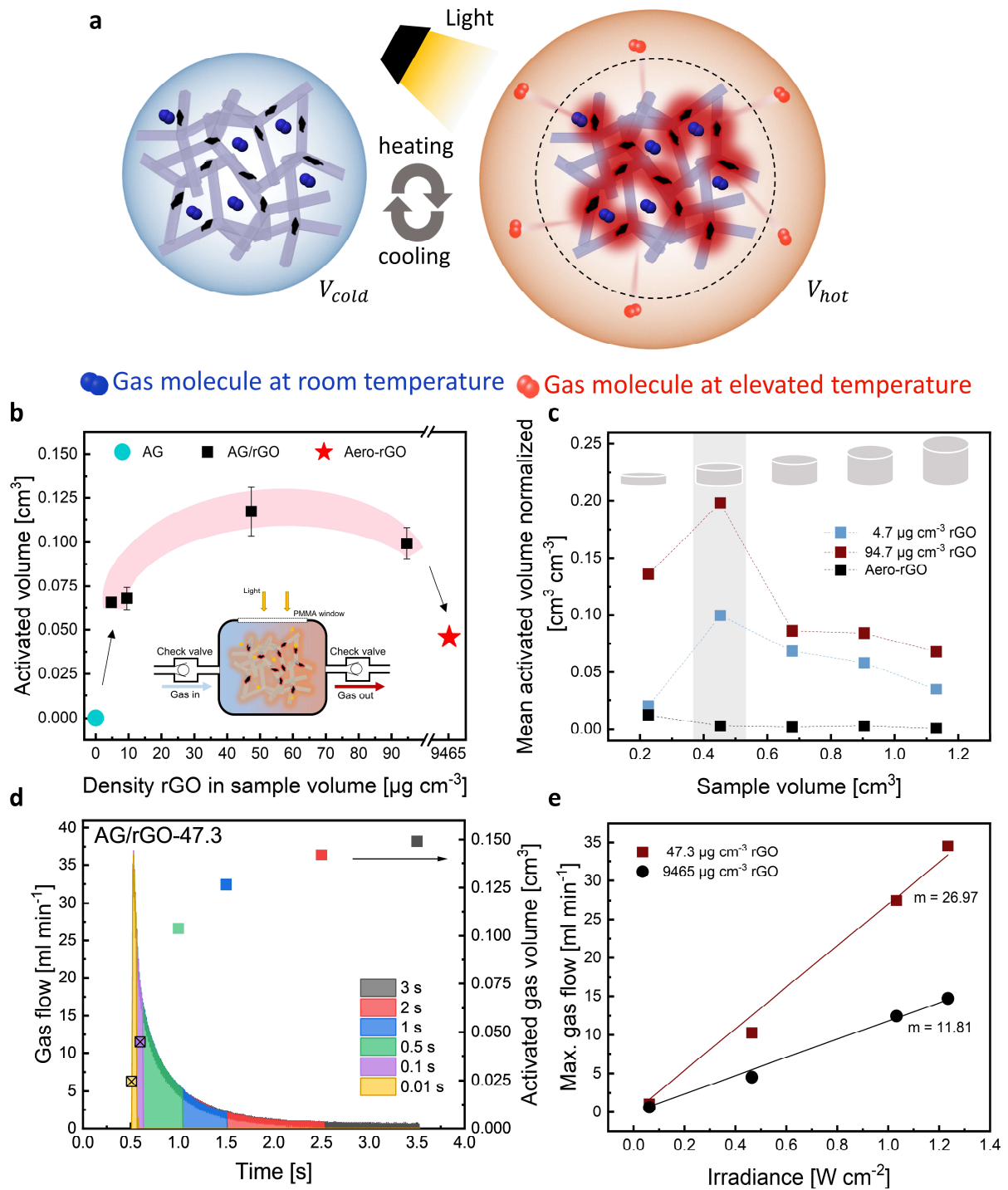

**Figure 4.** Characterization of volumetric gas activation. (a) Schematic of gas activation in AG/rGO hybrid aeromaterial. (b) Mean activated gas volume for 1 pulse (1 s) for pristine AG, AG/rGO with different volumetric loadings of rGO and pristine Aero-rGO. The inset shows a schematic of the measurement setup. (c) Mean activated volume normalized to sample volume measured for different sample heights of AG/rGO with low and high loading of rGO as well as for Aero-rGO. (d) Gas flow induced by gas activation versus time for different pulse lengths

with calculated activated gas volume on the second axis for a AG/rGO-47.3 sample. (e) Maximum detected gas flow versus irradiance for AG/rGO-47.3 and Aero-rGO.

**Figure 4c** shows the mean volume of activated gas normalized to the total volume of the hybrid aeromaterial for cylindrical sample geometries with varying height (2 mm - 10 mm) for the highest (AG/rGO-94.7) and lowest rGO loading (AG/rGO-4.7) (see **Figure S7**), as well as for pristine Aero-rGO. The highest gas activation for 1 s illumination time is reached by AG/rGO-94.7 with a height of 4 mm. Around ~ 20 % of the total aeromaterial volume is activated, which is by a factor of 66 higher compared to pristine Aero-rGO. The normalized activated gas volume decreases with further increasing height for both AG/rGO-94.7 and AG/rGO-4.7. In contrast, pristine Aero-rGO shows an almost constant value of ~0.3 % of volumetric gas activation irrespective of height. This significantly indicates that the amount of activated gas for pristine Aero-rGO is solely attributed to the afford mentioned surface effect (see also **Figure S8**). As an approximation, an ideal flat 2D surface with a radius of 6 mm provides a surface area of 1.13 cm$^2$ that can be used for gas activation by the photothermal effect, whereas for a cylindrical hybrid aeromaterial, e.g. AG/rGO-94.7, with radius 6 mm and height 4 mm provides a total surface area of ~500 cm$^2$ (details on the calculation can be found in SI Note 6). Assuming that the entire volume of the aeromaterial transducer is used for gas activation, the available surface area is by a factor of ~445 larger, enabling the rapid activation of larger gas volumes. Note, that due to the low density of gases, only the gas volume which is in close proximity to a heated surface can be rapidly activated.

    **Figure 4d** displays the gas flow generated by an AG/rGO-47.3 hybrid aeromaterial (d=12 mm and h=10 mm), which was illuminated with single light pulses as a function of pulse widths (0.01 s to 3 s) at a constant irradiance of 1.24 W cm$^{-2}$ (**Figure S9** for AG/rGO-94.7). The gas flow directly peaks after only 22 ms of illumination with a maximum flow of ~ 36 ml min$^{-1}$, before decreasing exponentially. The maximum detected gas flow for all pulse lengths was 36 ml min$^{-1}$ and the shape of the curve is the same for all applied pulse widths while the gas flow drops to zero immediately after illumination is turned off. As shown in **Figure 4d**, the activated gas volume is increasing with increasing pulse width. However, a maximum value for long pulse lengths (1 s - 3 s) of roughly 0.15 cm$^3$ is approached, as a mean temperature is reached and hence, no further gas activation occurs.

    **Figure 4e** shows the influence of light intensity on the maximum gas flow for Aero-rGO and AG/rGO-47.3 for a pulse length of 1 s. The maximum gas flow, as indirect measure for the activated gas volume, increases linearly with light intensity for both systems, as a higher

light intensity results in higher temperatures generated by the photothermal effect and hence, more gas is activated. However, the slope for AG/rGO-47.3 (26.97 ml cm$^2$ min$^{-1}$ W$^{-1}$) is more than twice as steep as for Aero-rGO (11.8 ml cm$^2$ min$^{-1}$ W$^{-1}$). While for Aero-rGO all light is absorbed close to the illumination side, an increase in light intensity only increases the surface temperature, which in return does only lead to a marginal change in activated gas volume. However, for the AG/rGO hybrid aeromaterial, higher light intensities also result in deeper light penetration, thereby resulting in a stronger increase of activated gas with increasing light intensity compared to pristine Aero-rGO.

Optimizing the volumetric light-to-heat conversion in AG/rGO hybrid aeromaterials requires the balance between light scattering, transmission and absorbance of light by photothermal material. The results indicate the importance to create a homogeneous heating effect in the entire volume rather than heating only parts of the structure to high temperatures in order to achieve highest values for gas activation within the structure. For the here shown volumetric loadings of rGO, AG/rGO-47.3 showed the best performance in terms of a volumetric heating effect and gas activation, as highest values for flow rate (36 ml min$^{-1}$) and activated gas volume (~0.15 cm$^3$) were measured, matching the results for temperature measurement, where the highest backside temperature for AG/rGO-47.3 was measured, and hence, indicating that light can interact with photothermal agents in a large part of the structure.

## 2.4 Applications

The observed enhancement in volumetric light–matter interaction provided by our optimized photothermal aeromaterial transducer, enables manifold wireless applications that make use of the rapidly activated gas volumes at high repetition rates. We here present first innovative proof-of-concept demonstrators for untethered, light-driven and light-controlled pneumatic soft actuators and wireless light-powered microfluidic pumps. Conventional soft actuators for soft robotic applications are based on pneumatic actuation systems that require pressurized air to create movements, which is usually delivered via air hoses.[22,23] This tethered connection strongly limits the autonomy, range of movement and the operational flexibility of soft robots.[22,23] While approaches to enable autonomous soft robotic system often rely on bulky carry-on systems[22,24], e.g. gas storage tanks[25], battery packs[22] or chemical reaction chambers[23], that increase weight and thus limit the operational time[33–35], we here demonstrate an autonomous untethered soft and lightweight pneumatic actuator that is purely powered and controlled by light. The here presented proof-of-concept demonstrators are operated using the best performing sample type AG/rGO-47.3.

The concept is based on the encapsulation of the photothermal aeromaterial transducer in a closed chamber that is directly connected to a conventional soft pneumatic actuator using a check valve (see **Figure S10**). Illuminating the transducer material with light, results in an increase in pressure in a closed system. **Figure 5a** shows the pressure rise measured for AG/rGO-47.3 as a function of light pulses for three different illumination frequencies and pulse-widths. As can be seen, in **Figure 5a**, the pressure can be adjusted by the number of light pulses. A maximum increase in pressure of ~87 mbar is reached in ~40s, corresponding to 10 light pulses (illumination time of 1 s with a recovery time of 3 s). Thereby, a pneumatic soft actuator can be deformed and controlled by light, as shown in **Figure 5b**. Each light pulse creates an increase in pressure, bending the soft robotic arm in a step-like motion until saturation pressure is reached. Mounting the pneumatic soft actuator directly to the chamber without any check valves even allows to bend and relax the soft pneumatic actuator with only one single light pulse in a reversible manner (see Supplementary Video 2).

Next to applications in soft robotics, the aeromaterial transducer can be utilized for wireless, light-powered and light-controlled fluidic (gas and liquid) micro-pumps. **Figure 5c** shows the maximum gas flow over 77 light pulses (pulse width of 1 s at a frequency of 0.5 Hz and an irradiance of 1.24 W cm$^{-2}$) for AG/rGO aeromaterial transducers with different volumetric rGO loadings as well as pristine Aero-rGO. AG/rGO-47.3 shows the highest measured gas flow rate of ~ 36 ml min$^{-1}$, which is by more than 33 % higher compared to the pristine Aero-rGO. Furthermore, while all AG/rGO hybrid aeromaterials show an almost constant gas flow over the entire cycle range, the gas flow for Aero-rGO decreases slowly from ~14 ml min$^{-1}$ until a constant value of ~10.8 ml min$^{-1}$ is reached after roughly 45 cycles. We suspect, that the decrease is associated with a heat accumulation at the illuminated surface. Thereby the difference in temperature ($T_{light\_on}$ vs $T_{light\_off}$) decreases over time, resulting a lower volume expansion until an equilibrium state is reached.

The pumped fluidic volume can be easily controlled by varying pulse lengths and applied light frequencies, as demonstrated in **Figure 5d, e,** showing a light driven micro-pump with a pump rate of ~ 3.51 ml min$^{-1}$ pushing a droplet of dyed water through a glass tube (see Supplementary Video 1). Furthermore, a wireless light-powered and light-controlled membrane pump could be realized (see **Figure 5f-h** and Supplementary Video 3), utilizing the rapid switching in pressure to repeatable strain a flexible membrane in order to pump a liquid (details on the set-up can be found in **Figure S11**), achieving a pump rate of ~ 0.44 ml min$^{-1}$ without optimization.

It should be noted, that for all the here suggested application scenarios, the lightweight nature of the hybrid aeromaterials makes them extremely resistant to vibrations and forces, e.g. caused by high accelerations, ensuring that no structural damage occurs over time, especially when encapsulated.

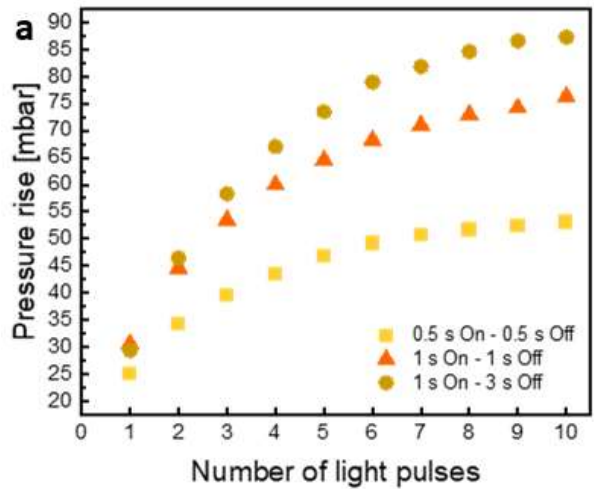
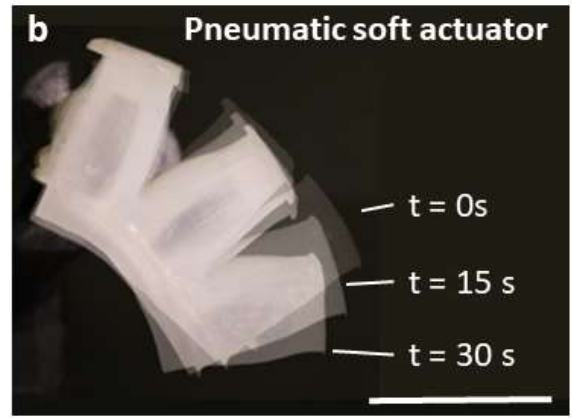
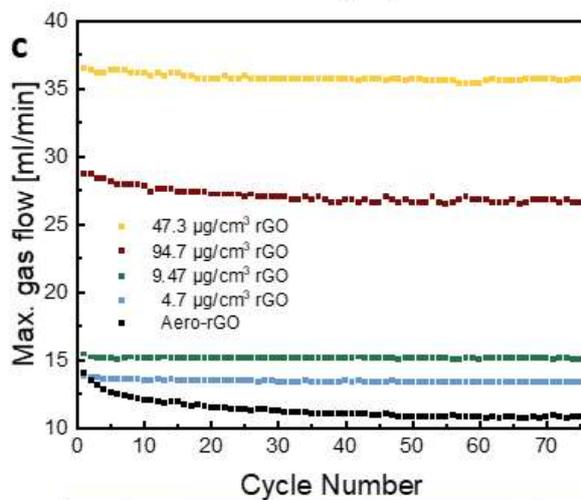
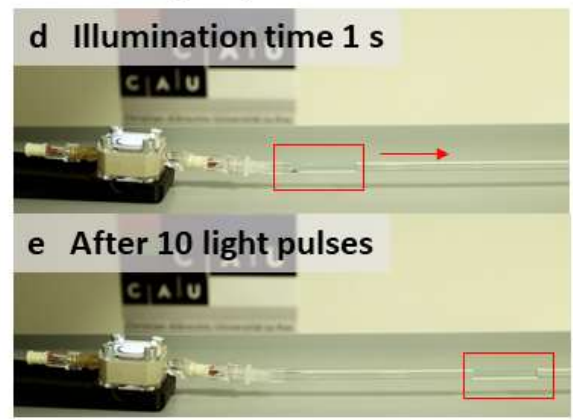
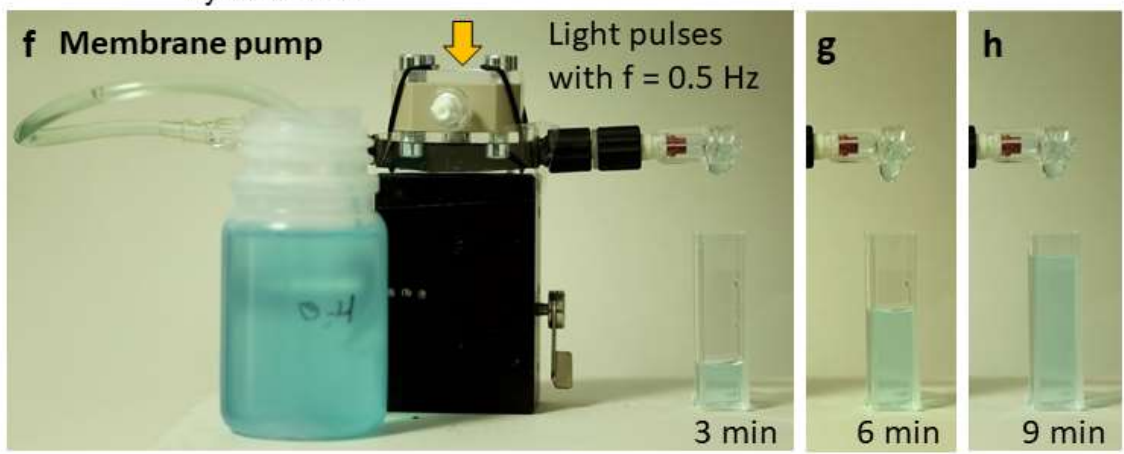

**Figure 5** Application scenarios for micro- and nano-engineered AG/rGO hybrid aeromaterials. (a) Pressure rise generated by an AG/rGO-47.3 sample in an external pressure chamber versus number of applied light pulses for different illumination frequencies. (b) Merged photograph of free standing (untethered) soft pneumatic actuator powered by light at three different time points of actuation. (c) Maximum detected flow versus cycle number for cyclic illumination with 0.5 Hz and an irradiance of 1.24 W cm$^{-2}$. (d),(e) Photograph of micro-pump pushing a droplet of dyed water through a glass tube, showing the covered distance after 10 light pulses. (f)-(h) Sequence of images of a wireless light-powered membrane pump in operation after 3 min, 6 min and 9 min.

## 3. Conclusion

In conclusion, we here show a novel fabrication concept for the controlled combination of different nanomaterials and their functionalities in a hybrid 3D assembly, thereby overcoming the limitations of their pristine variants. We demonstrate a hybrid $SiO_2$/rGO photothermal aeromaterial transducer that enables an ultra-rapid and enhanced volumetric photothermal response by a factor of 2.44 compared to the pristine carbon nanomaterial-based variant. By combining the light scattering properties of a highly porous $SiO_2$ framework structure with the photothermal properties of rGO we are able to tailor the light–matter interaction volume in a controlled manner by adjusting the volumetric density of rGO nanoflakes within the hybrid aeromaterial. Utilizing the enhanced photothermal interaction volume in combination with the high surface area and low volumetric heat capacity of the porous transducer material enables an ultra-rapid activation of macroscopic gas volumes that opens up new functionalities and application scenarios based on the conversion of light into heat. While we here demonstrate first innovative proof-of-concept demonstrators for untethered, light-driven and light-controlled pneumatic soft actuators and wireless light-powered microfluidic pumps, the hybrid aeromaterials could be used to generate a directional pressure wave, e.g. for light-controlled, directional motion. In perspective, a directional and also dynamic illumination of different sections of an extended hybrid aeromaterial could generate several pressure waves emanating from parts of the material, which result in a certain direction of propagation through destructive and constructive interference. By controlling the frequency and changing the order of illumination, the resulting pressure wave could be shaped in relatively arbitrarily manner. Next to that, the concept paves the way for advances in the fields of environmental remediation, such as the synthesis of solar fuels by gas phase (photo-)catalysis, solar thermal heating and desalination, as well as photoacoustics. For example, it is conceivable to use the photothermal transducer material for photocatalysis applications to simultaneously pump gas and trigger a chemical reaction by temperature activation. Expanding the concept to fabricate lightweight hybrid aeromaterial transducers by utilizing the unique properties of other 1D and 2D nanomaterials, including quantum dots, MXenes and catalysts, opens up new functionalities well-beyond that of photothermal conversion.

## 4. Materials and Methods

**Materials**: Polyvinyl butyral, zinc (grain size 1-5 μm), ethanol, TEOS, ammonia hydroxide solution, graphene oxide dispersion, L-ascorbic acid

**Graphene oxide** (GO) was obtained by procedures reported elsewhere[26]. Dispersion were prepared by tip sonication in water.

**Aeroglass** structures were fabricated using tetrapodal ZnO (t-ZnO) as sacrificial template for a wet chemical coating process with $SiO_2$ based on the Stoeber synthesis[27]. The fabrication of sacrificial templates is described elsewhere[17,28,29]. Briefly, t-ZnO was produced via a flame transport synthesis[29,30], pressed into desired shape using a metal mold, and subsequently sintered in an oven (1150°C for 5 hours) to create an interconnected network of ZnO tetrapods. The wet chemical approach for coating the ZnO network with a thin layer of $SiO_2$ is based on the Stoeber process. In detail, two solution with different ratios of ethanol, tetraethylorthosilicate (TEOS) and ammonia hydroxide solution (25%), namely V1 (ratio 10:0.1:3) and V2 (ratio 10:0.05:3) were used in this work. The silica solution was prepared by mixing ethanol and TEOS, followed by the addition of ammonia hydroxide solution. The sacrificial ZnO templates were immersed in the silica solution for 45 min during which the solution turned turbid and the entire ZnO template was coated with a thin layer of amorphous $SiO_2$. The coating was performed in a two-step process, first a coating with ratio V1 was applied, followed by rinsing in ethanol and drying on a Teflon plate, before the second coating was applied using ratio V2. Samples were stored in distilled water for 24 h, followed by wet chemical removal of the sacrificial ZnO template with 1M hydrochloric acid. To remove residuals, the samples were washed thoroughly with water, before being transferred into pristine ethanol for critical point drying (Leica EM CPD300).

**Functionalization of aeroglass**. Prior to the two-step silica coating step, the ZnO templates were infiltrated with a water-based dispersion of nanomaterial, as described elsewhere[17,28]. This led to a partial coating of the nanomaterial on the ZnO tetrapod arms. To control the amount of nanomaterial within the aeroglass structure the concentration of the nanomaterial dispersion was adjusted. For functionalization with reduced graphene oxide the ZnO templates were coated with graphene oxide flakes which were reduced in an additional step after the wet chemical removal of ZnO using diluted L-ascorbic acid (24 h, 50°C).

**SEM** measurements were performed using a Zeiss Supra 55VP.

**TEM** was carried out using a FEI Tecnai F-30 G2 Stwin operated at 300 kV. The microscope is equipped with an EDAX Si/Li detector for elemental analysis and a Gatan Tridiem 863P post column image filter (GIF) allowing for electron energy loss spectroscopy (EELS) measurements. AG and AG/rGO specimens were prepared for TEM investigation by grinding a small amount of material in butanol followed by drop coating onto lacey carbon coated Cu TEM grids.

**UV-vis spectroscopy** was carried out in water to avoid measurement artefacts due to light scattering of the samples. A sample thickness of 2 mm was chosen to ensure a transmitting beam through the sample. A customized photometer (light source HL-2000-FHSA-LL by Ocean Insight, Ostfildern, Germany and spectrometer Flame-S-VIS-NIR_ES by Ocean Insight, Ostfildern, Germany), controlled by a customized LabVIEW program (National Instruments, Austin, TX, USA, version 2012) was used for the measurement and the data was evaluated using Matlab (The MathWorks Inc., Natick, MA, USA, version R2021a).

**Raman measurements** were conducted using an alpha300 RA (WITec) microscope with a triple grating spectrometer (600 g mm$^{-1}$) and a charge-coupled device detector. The excitation wavelength of the laser was 532.2 nm with a laser spot size on the sample of ∼1.41 μm and a maximum power of 52 mW.

**Light source** for illuminating the samples with white light was a beamer (Acer, DLP Projector, DNX0906) in combination with an overhead projector lens to focus the beamer light to a spot size of 12 mm. A spectrum was recorded with an Ocean Optics spectrometer FLAME-S-XR1-ES (200-1025nm), see **Figure S12**. For controlling the illumination time, as well as the light power, a power point presentation with pre-set display times and varying grayscale of displayed images was used.

**Thermography** was performed placing the samples directly in the focus point of the light source and temperature was recorded with an IR camera (InfraTec, PIR uc 180) from three different directions. In more detail, the temperature of the sample frontside (illumination side) and the sample backside were recorded as well as the side view to gain information about the temperature gradient within the sample. The samples were cylindrical shaped with a diameter of 12 mm, height of 4 mm to 10 mm and a conical shaped taphole on the illumination side of the sample (only for comparison of different graphene loadings). The samples were illuminated for 1 s with an irradiance of 1.67 W cm$^{-2}$ and the image with maximum temperature was selected. The irradiance was measured with a power meter (Thorlabs Photometer, S425C-L with Interface PM100USB and Thorlabs Opitcal Power Monitor).

**Measurements on volumetric gas activation** were conducted using the same set-up as for thermography, but placing the samples in an enclosed chamber with a polymethylmethacrylate window on the illumination side and two check valves (see Figure S5 and Figure S6), from which one was connected to a glass tube which contained 1 ml of dyed water. The samples have been illuminated with a frequency of 0.5 Hz and an irradiance of 1.24 W cm$^{-2}$. In total, 10 light pulses were performed for each sample. The activated gas volume was calculated from the distance the water droplet was pushed through the glass tube and the diameter of the tube and averaged over 10 light pulses. The activated gas volume for varying sample heights was measured with spacers to ensure an equal distance to the light source as well as no free volume above the illumination side.

**Flow measurements** were conducted using the enclosed chamber equipped with two check valves and a flow sensor (Honeywell AWM3100V). Samples have been illuminated with an irradiance of 1.24 W cm$^{-2}$ and pulse widths ranging from 0.01 s up to 3 s. The activated gas volume was calculated by integration of the measurement data.

**Light transmission** was measured with a power meter (Thorlabs Photometer, S425C-L with Interface PM100USB and Thorlabs Opitcal Power Monitor) placed directly behind the sample which was illuminated with an irradiance of 1.24 W cm$^{-2}$ for 3 s.

**Pressure measurements** were conducted using a pressure sensor (Kistler, 4260A) mounted on an empty chamber which was connected via a check valve to the enclosed chamber containing the sample as described before.

## Declaration of Competing Interest

The authors declare no conflict of interest.

## Acknowledgment

The authors acknowledge funding from the European Union's Horizon 2020 Research and Innovation Programme under grant agreement No GrapheneCore3 881603 and from the German Research Foundation (DFG) under grant KI 1263/17-1.

## Author contributions

L.M.S., F.S., S.S. and R.A. developed the hybrid AG/rGO aeromaterials and designed the study. A.S.N. prepared the GO dispersion. L.M.S. fabricated the samples, carried out the experiments and evaluated the data. H.Q. performed Raman measurements. N.K. and L.K. performed TEM measurements and evaluated the TEM specific data. L.M.S., F.S. and R.A. finalized the study and wrote the paper. All of the authors have contributed to the discussion of the results and reviewed the manuscript.

**Supporting Information**

Supporting Information is available from the Wiley Online Library or from the author.

**Data availability**

The data that support the findings of this study are available from the corresponding authors upon request.

# Supplementary Information

**This PDF file includes:**

Supplementary Figures S1 to S13

Supplementary Notes 1 to 5

    Supplementary Note 1 - TEM investigation of aeroglass

    Supplementary Note 2 – Comment on UV-vis spectroscopy

    Supplementary Note 3 – Light scattering in aeroglass

    Supplementary Note 4 – Calculation of volumetric specific heat capacity

    Supplementary Note 5 – Calculation of specific surface area

Captions for Supplementary Videos 1 to 3

References

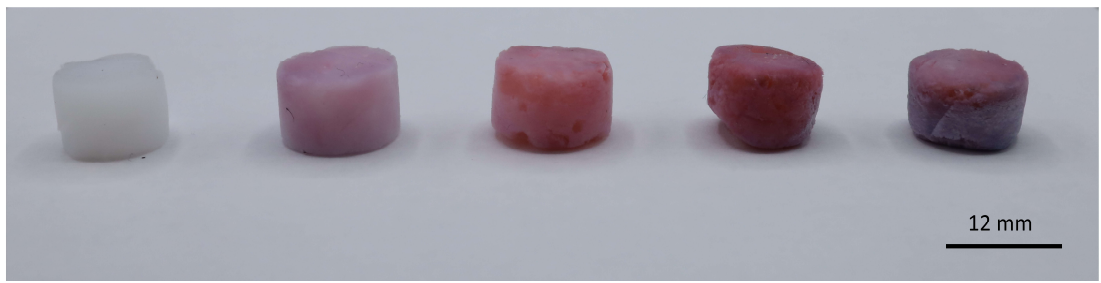

**Figure S1.** Aeroglass (AG) functionalized with increasing concentration (from left to right) of gold nanoparticles.

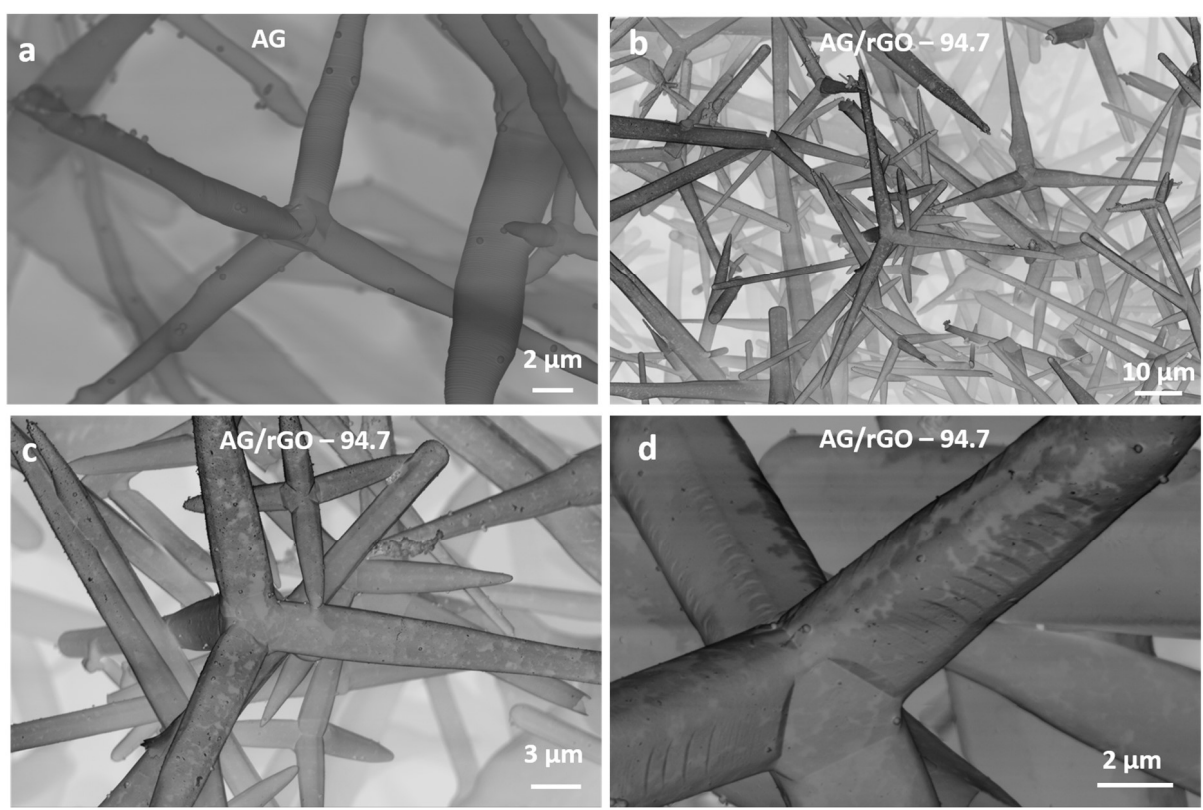

**Figure S2.** Detailed scanning electron microscopy characterization of (a) pristine aeroglass (AG), (b-d) AG/rGO with volumetric loading of 94.7 µg cm$^{-3}$ rGO with increasing magnification. The stained surface of the SiO$_2$ microtubes indicates the homogeneous coating with rGO flakes.

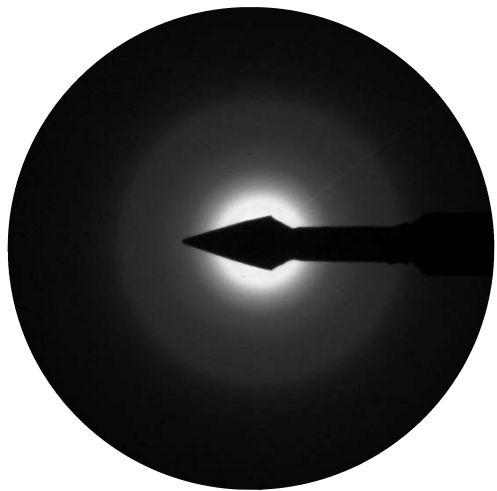

**Figure S3** Diffraction pattern recorded by TEM of pristine aeroglass revealing the amorphous character of $SiO_2$ microtubes.

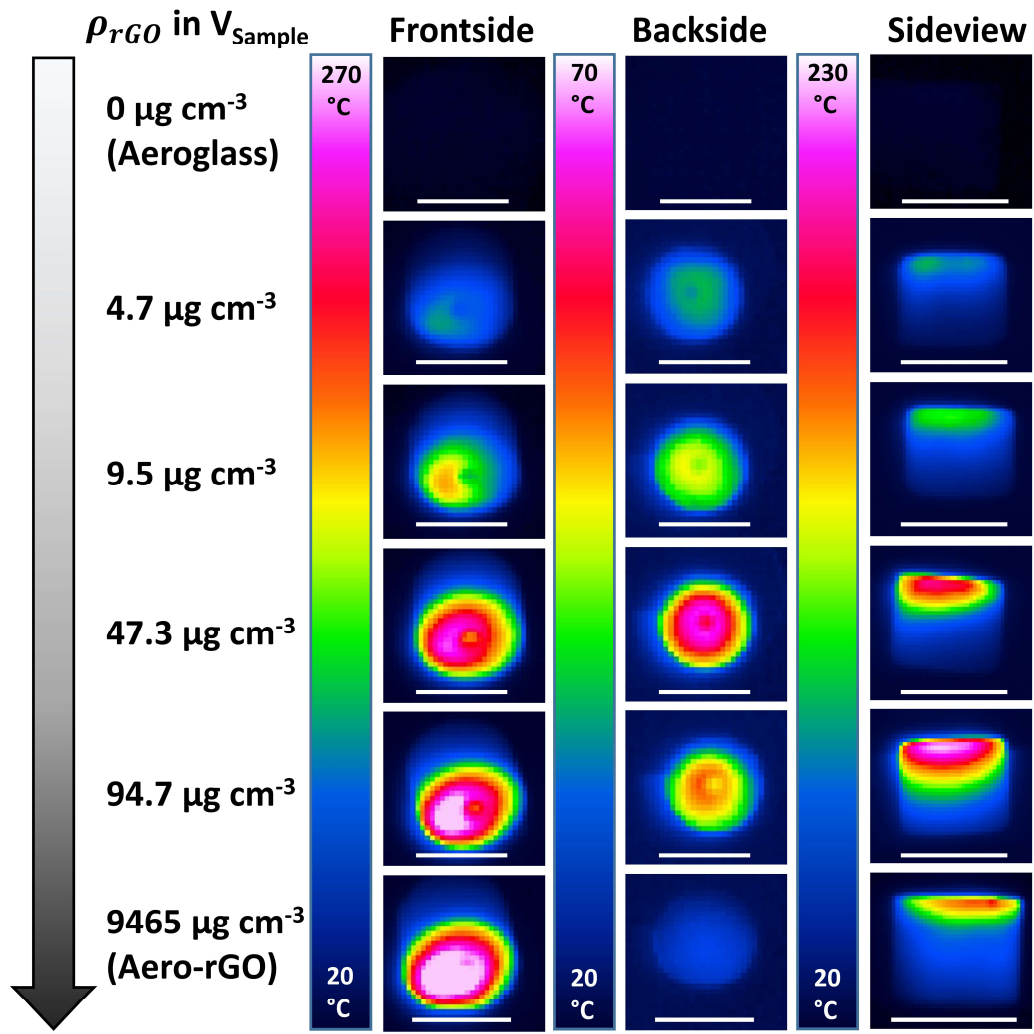

**Figure S 4.** Infrared images of hybrid aeroglass with increasing volumetric rGO loading from top to bottom. Images were recorded from the fontside (illumination side), backside, and the sideview of the samples for an illumination for 1 s with an irradiance of 1.67 W cm$^{-2}$. Scale bar is 12 mm.

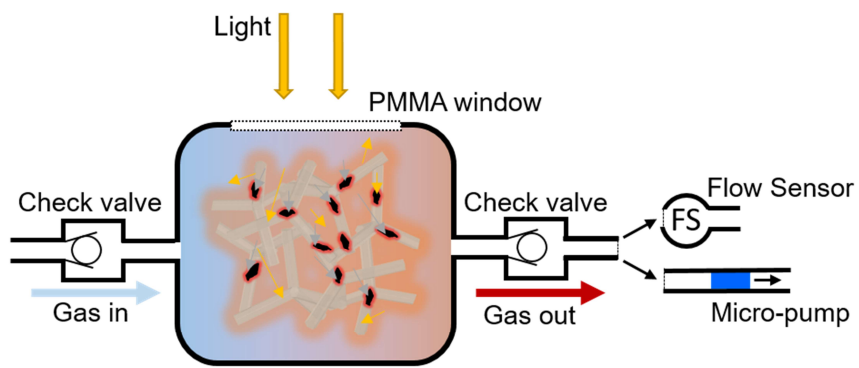

**Figure S5.** Schematic of air-tight chamber with multiple connection options.

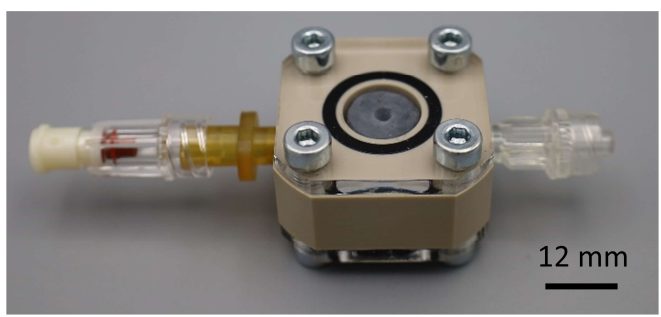

**Figure S6.** Photograph of air-tight chamber containing a AG/rGO sample.

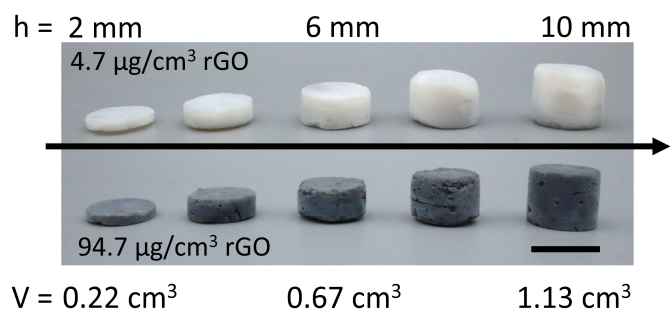

**Figure S7.** Photograph of AG/rGO-4.7 and AG/rGO-94.7 with varying sample dimensions. Scale bar is 12 mm.

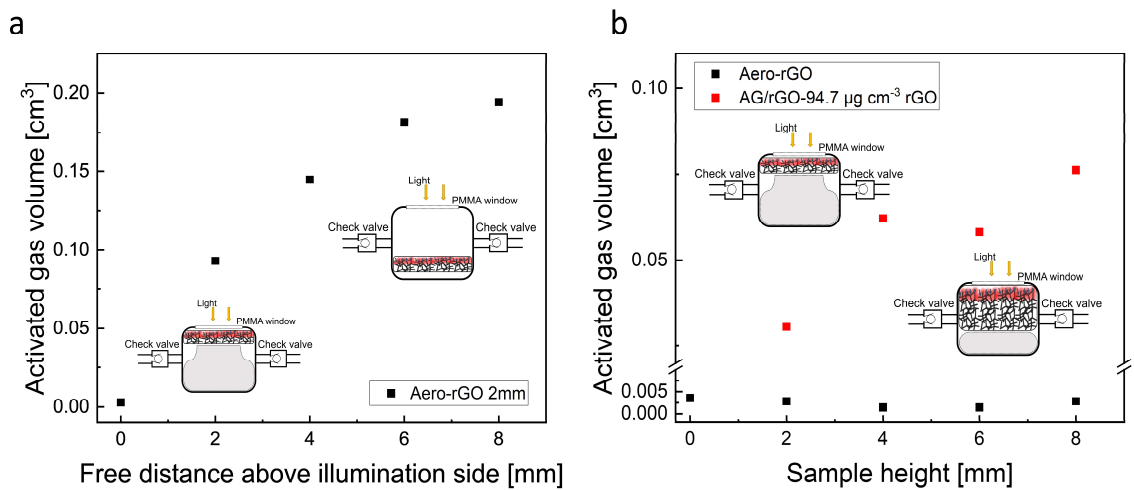

**Figure S8.** (a) Activated gas volume of (a) Aero-rGO with 2mm sample height. The sample was placed in an air-tight chamber whereby the distance to the polymethylmethacrylate window was varied from 0 mm to 8mm. This resulted in varying free volumes above the illumination side, as demonstrated with the schematic insets. With increasing free volume above the illumination side, the activated gas saturates. (b) Activated gas volume for Aero-rGO and AG/rGO-94.7 with increasing sample height and a distance of 0 mm to the polymethylmethacrylate window. In this configuration only the volumetric gas activation by the photothermal effect is measured, demonstrating that only the surface region contributes to the gas activation. The insets show the position of the samples in the air-tight chamber and the heated surface region for Aero-rGO. In contrast, AG/rGO-94.7 shows an increase in gas volume with sample height, indicating a volumetric photothermal conversion.

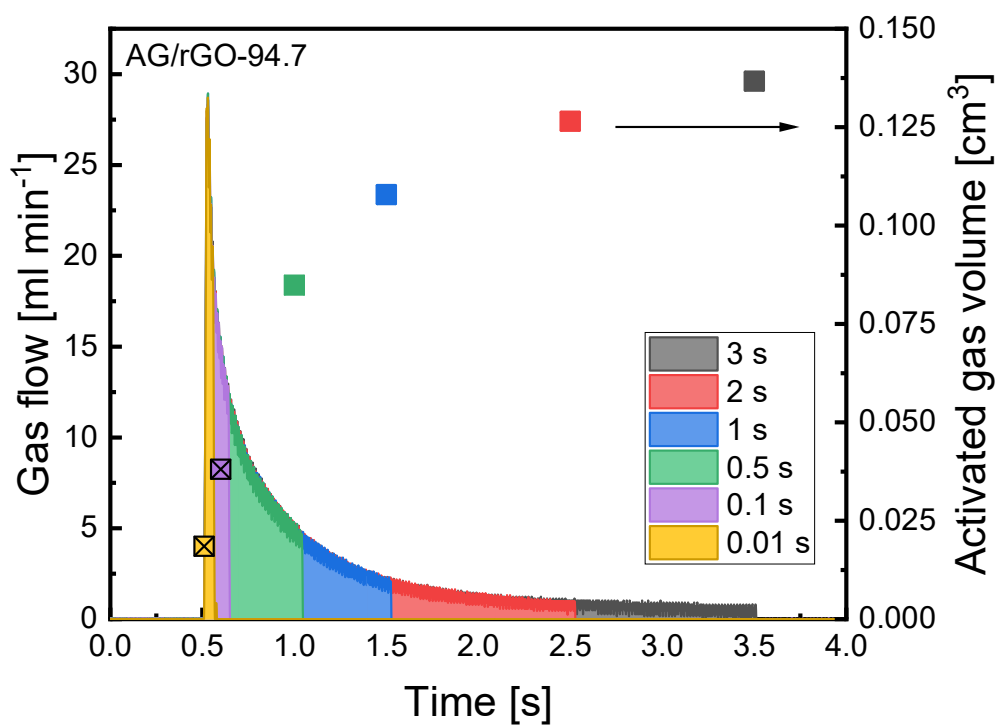

**Figure S9.** Gas flow generated by AG/rGO-94.7 illuminated with varying times ranging between 0.01 s and 3 s.

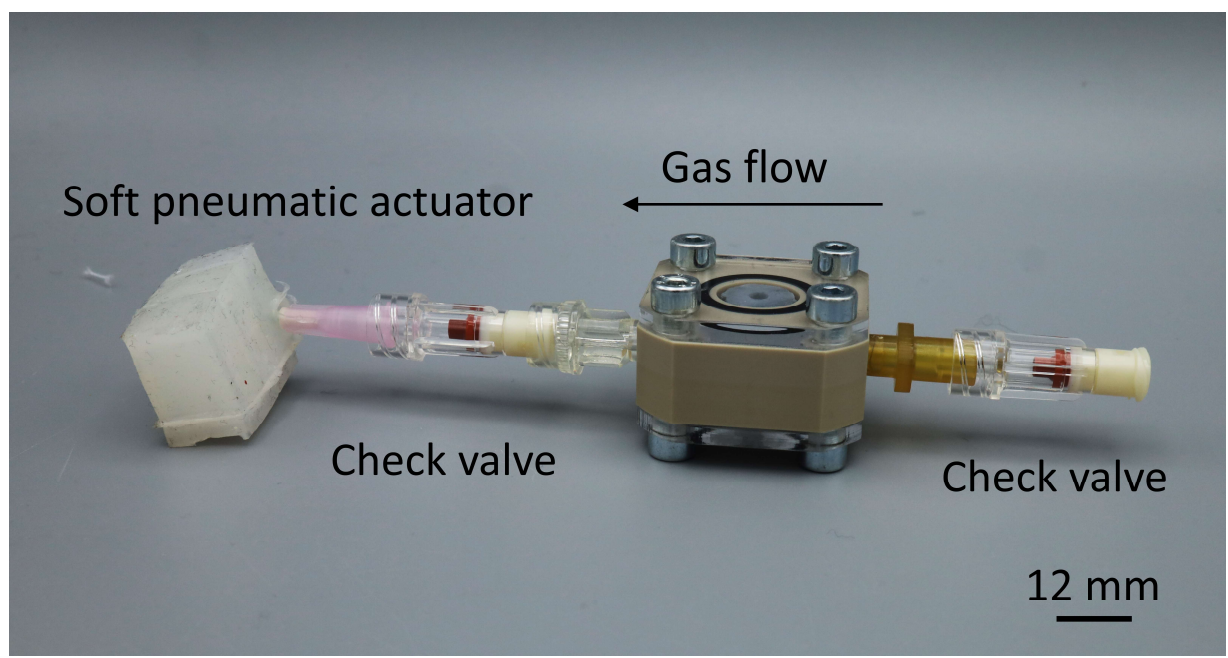

**Figure S10.** Set-up for untethered soft pneumatic actuator that can be controlled by light. The check valves allow a directional gas flow upon illumination of the aeromaterial transducer resulting in a step-wise bending of the soft pneumatic actuator.

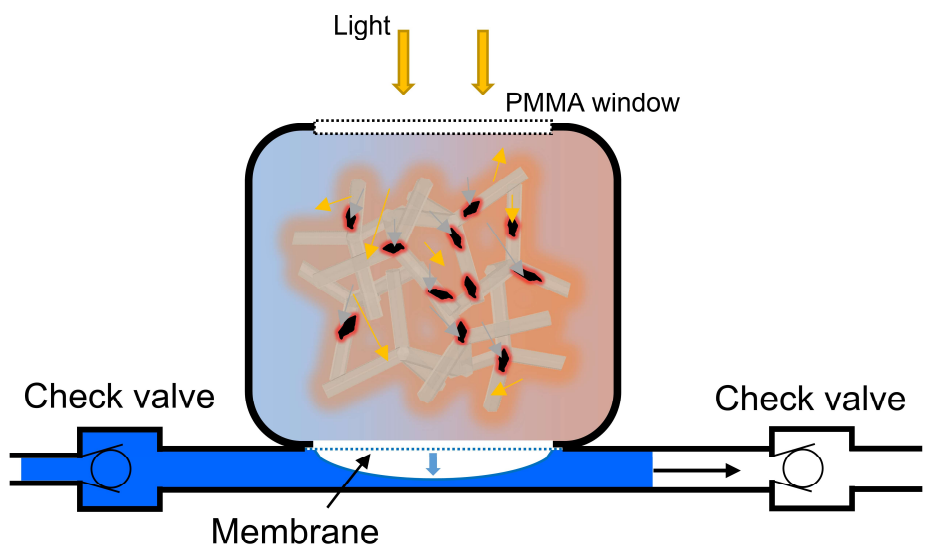

**Figure S11.** Schematic concept of wireless light-powered membrane pump. Illumination of AG/rGO in an enclosed chamber results in a pressure rise that causes a repeatable straining of the flexible membrane.

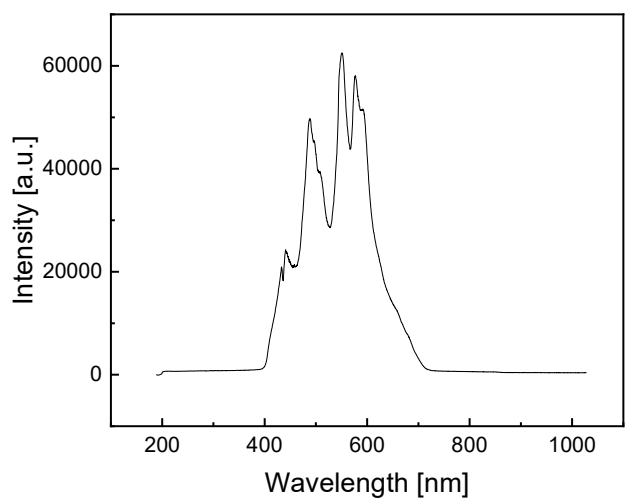

**Figure S 12.** Spectrum of beamer (Acer, DLP Projector, DNX0906).

**Supplementary Note 1 - TEM investigation of aeroglass**

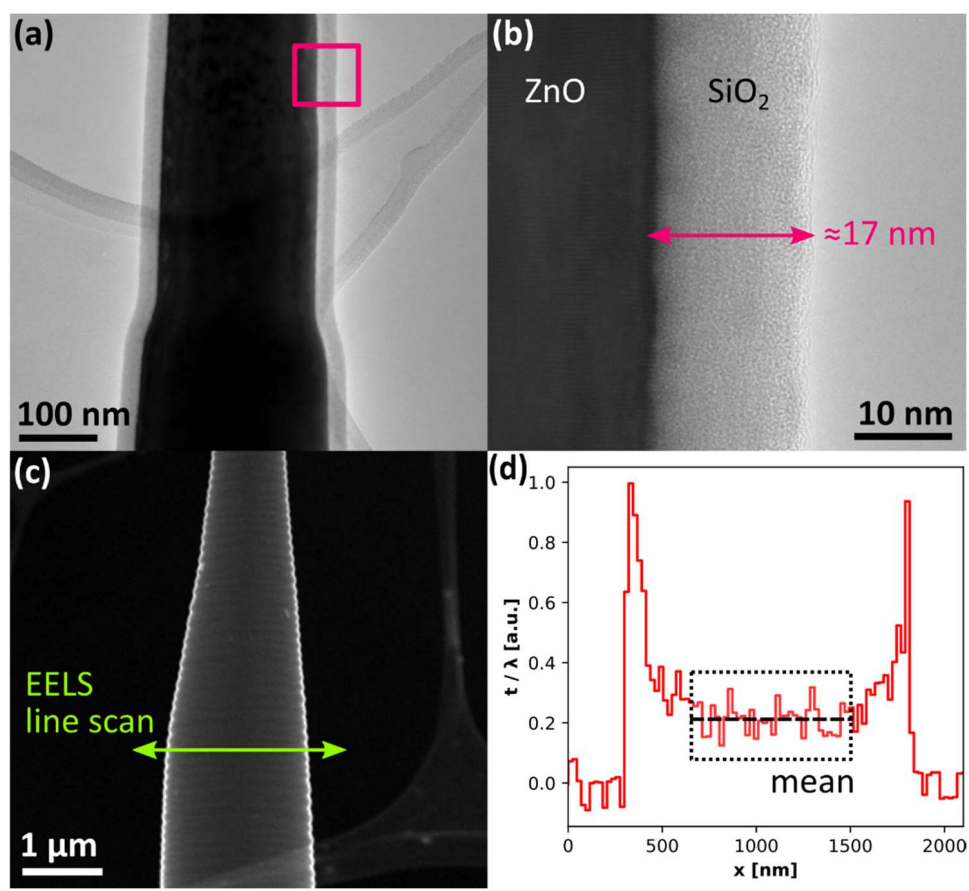

**Figure S13.** (a) TEM image of a ZnO tetrapod arm coated with a thin $SiO_2$ layer under identical parameters as for AG fabrication showing a homogenous layer thickness. (b) HRTEM micrograph of the area marked in (a). The $SiO_2$ layer can clearly be distinguished from the ZnO due to mass thickness and diffraction contrast. (c) HAADF-STEM image of an AG microtube. (d) Relative thickness profile of the microtube from (c) along the indicated line scan path. The (relative) wall thickness of the tube is determined by taking the mean from the indicated area.

TEM investigation of AG and AG/rGO was carried out both to reveal its general properties and functionalization (as shown in the main manuscript) as well as to determine the $SiO_2$ wall thickness. In order to determine the $SiO_2$ wall thickness two complementary approaches can be used. First, the wall thickness can be determined directly by fabricating specimens that are coated with $SiO_2$ in an identical process used for AG synthesis but without removing the ZnO. Accordingly, ZnO tetrapods coated with a $SiO_2$ layer of identical thickness to the corresponding AG wall thickness are obtained, see **Figure S13a**. The $SiO_2$ layer thickness can then directly be determined using HRTEM due to the strong contrast between the crystalline ZnO and the amorphous $SiO_2$, see **Figure S13b**. By investigating multiple tetrapod arms and locations an average layer thickness of 17 +- 0.5 nm is found. The uncertainty is given by the standard error of the measurements.

The second approach can be applied to AG directly. By utilizing the electron energy loss spectroscopy (EELS) log ratio method[1] the relative thickness in electron beam direction can be determined in multiples of the inelastic electron mean free path (IEMFP). Using appropriate approximations for the IEMFP [1,2], the absolute thickness can then be subsequently obtained. If the EELS log ratio method is combined with STEM EELS, localized thickness information can be gathered. Performing line scans across a hollow AG microtube as shown in **Figure S13c**, then yields the thickness profile across the AG microtube given in **Figure S13d**. Taking the mean of the center area of approximately constant layer thickness, due to the curvature of the tube, and dividing by a factor of two taking into account that the electron beam passes through the $SiO_2$ layer twice, yields the relative layer thickness. Measuring multiple tubes and positions results in a relative AG wall thickness of 0.11 +- 0.025 IEMFP. Using the approximation given by Malis et al.[1] results in an IEMFP of 143.5 nm while the formula from Iakoubovskii et al.[2] using a mass density of 2.0 g cm$^{-3}$ for stoeber silica [3] results in an IEMFP of 197 nm. Multiplying the relative thickness determined via EELS log ratio with the calculated IEMFPS for (stoeber) $SiO_2$ yield the absolute wall thickness of 15.8 +- 0.4 nm using Malis et al.[1] and 21.7 +- 0.5 nm using [2,3]. The deviation between the two thickness values denotes a typical shortcoming of the EELS log ratio method where the accuracy of absolute thicknesses is dominated by the uncertainty in the IEMFP, which is for most cases not precisely known. In this investigation, however, the thickness value of 15.8 +- 0.4 nm determined via the EELS log ratio method using the IEMFP calculation from Malis et al.[1] is remarkably close to the directly determined layer thickness of 17 +- 0.5 nm. If multiple specimens of AG or different layer thickness were to be measured the direct thickness value could be used to calibrate the EELS log ratio measurements using an effective experimental IEMFP. In this case an IEMFP of 154.5 nm would correspond to the directly determined layer thickness.

The observations and conclusions require that the $SiO_2$ layer thickness is not changed during ZnO removal in the AG fabrication process. Both the chemical inertness of $SiO_2$ towards the etching agent as well as the closely matching thickness values support this assumption.

**Supplementary Note 2 – Comment on UV-vis spectroscopy**

UV-vis spectroscopy was carried out for pristine AG, AG/rGO with varying volumetric loading of rGO (4.7 µg cm$^{-3}$ rGO, 9.47 µg cm$^{-3}$ rGO, 47.3 µg cm$^{-3}$ rGO and 94.7 µg cm$^{-3}$ rGO), see **Figure 2f** in main manuscript, as well as for pristine Aero-rGO (9465 µg cm$^{-3}$ rGO), see **Figure S14**. The measurements were conducted in water to avoid measurement artefacts by strong light scattering. The index of refraction of water is close to that of SiO$_2$, therefore almost no light scattering occurs. In contrast, when measuring aeroglass structures in air, nearly no light would reach the detector, as the samples scatter light to all space angles. Pristine AG shows low absorbance for 400 nm to 500 nm with a decreasing trend for higher wavelengths. It should be noted, that although the measurements were conducted in water to avoid scattering, this might still cause a deviation of the absorbance values. AG/rGO-4.7 shows a slightly higher absorbance over the measured wavelength range compared to pristine AG. In contrast, AG/rGO-94.7 rGO shows a strongly enhanced absorbance over the entire wavelength range with highest absorbance for 450 nm to 550 nm and a decreasing trend towards higher wavelengths and the near infrared range (NIR). The absorbance of AG/rGO with intermediate degree of functionalization shows the same trend with absorbance values in between the values for low and high loadings of rGO. For comparison, the absorbance of pristine Aero-rGO was measured as well. The result is shown in **Figure S14**. All light is directly absorbed and almost no light reaching the detector.

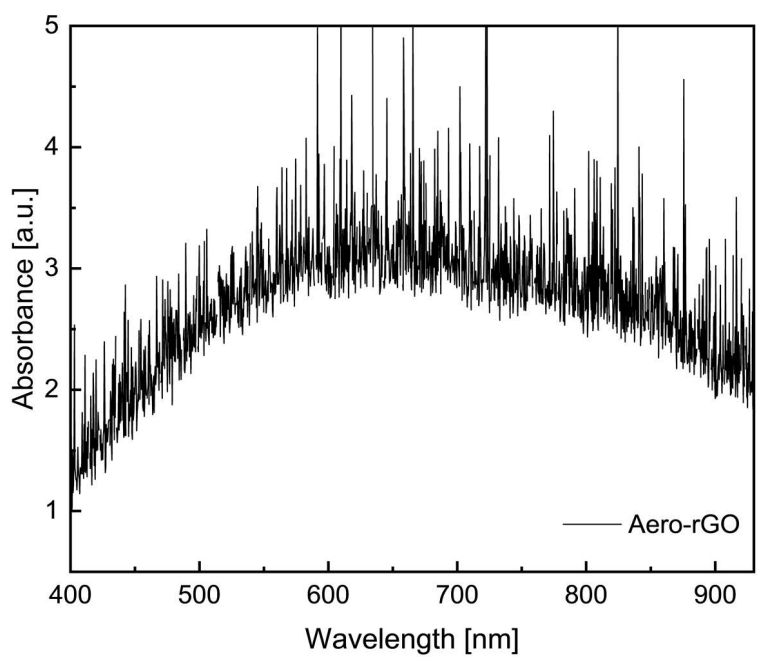

**Figure S14.** UV-vis absorption spectra of Aero-rGO.

**Supplementary Note 3 – Light scattering in aeroglass**

Pristine AG (**Figure S15a**) shows a unique open porous hierarchical structure with different sized features such as microsized voids (100-300 µm), which are much larger than the wavelength of visible light, the diameter of the microtubes (1-3 µm), which is in the same order of magnitude as the wavelength of visible light, and the nanoscopic wall thickness, which is much smaller than the wavelength of visible light, as schematically shown in **Figure S15b,** thus, creating an optical disorder system in which the nanoscopic wall thickness provides scattering centers for Rayleigh type light scattering.[4] Scattering occurs randomly in any space direction. Light impinging on the structure can be scattered multiple times before leaving the structure, thus, acting as a diffuser for light, as shown in **Figure S15c**.

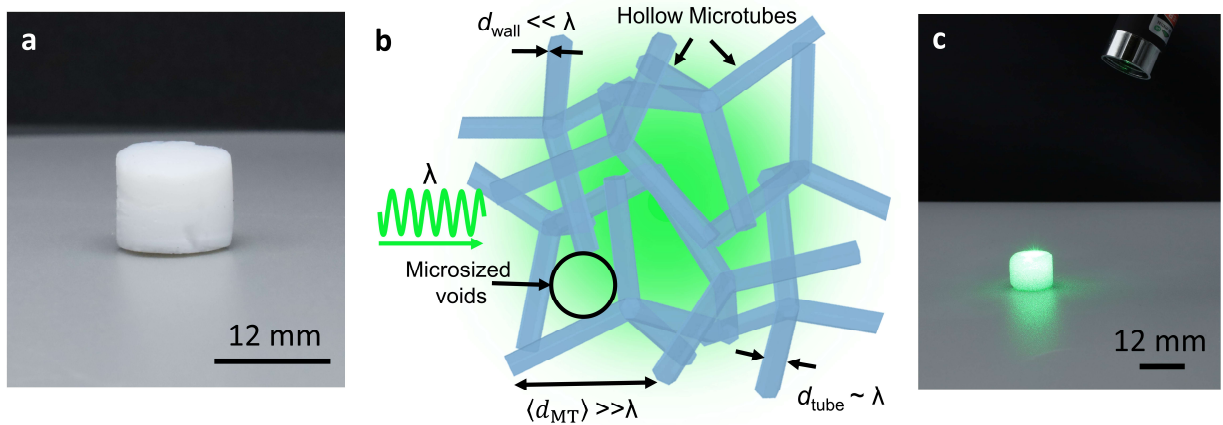

**Figure S15.** (a) Photograph of pristine AG showing optically white appearance. (b) Schematic of microstructure composed of hollow interconnected microtubes with hierarchical features sizes. (c) Photograph of AG illuminated with a green laser, scattering the light in any space direction.

## Supplementary Note 4 – Calculation of volumetric specific heat capacity

The volumetric specific heat capacity of highly porous and low density materials is difficult to determine with standard methods as e.g. the laser flash method. Therefore, the volumetric specific heat capacity was calculated as a rough estimate based on the molar specific heat capacity of amorphous $SiO_2$: $c \approx 50\ J\ mol^{-1}\ K^{-1}$ [5]

The specific heat capacity of $SiO_2$ is calculated by

$$c_p = c * M, \qquad (S1)$$

with M being the molar mass of $SiO_2$. As the molar mass changes for different crystallinities, we here assume a value of $M = 60.1\ g\ mol^{-1}$, hence, $c_p = 0.83\ J\ g^{-1}\ K^{-1}$.

The volumetric heat capacity s of a material results from:

$$s = c * \rho, \qquad (S2)$$

with $\rho$ being the density of the material. With a density of $\approx 3\ mg\ cm^{-3}$ the volumetric heat capacity is calculated as $2.5\ kJ\ m^{-3}\ K^{-1}$.

## Supplementary Note 5 – Calculation of specific surface area

The high specific surface area of aeroglass is crucial for the heat transfer to the gas phase. Due to the lightweight and highly porous macroscopic structure standard techniques to determine the gravimetric surface area, such as BET, are difficult to apply, because the overall amount of material per volume is very low. Therefore, the specific surface area is estimated by the following calculation.

As the fabrication process is based on a template based approach, first the volumetric surface area (VSA) of the sacrificial tetrapodal ZnO (t-ZnO) template is calculated. For this, it was assumed that the ZnO tetrapods are composed of 4 rods with a diameter d = 2 µm and a length l of 25 µm.

The surface area of a single ZnO tetrapod $A_{tetrapod}$ is calculated by:

$$A_{tetrapod} = 4 * \frac{2\pi * \frac{d}{2} * \left(\frac{d}{2} + l\right)}{10^8} \qquad (S3)$$

Knowing the template geometries, the density of the template ($\rho_{template} = 0.3\ g\ cm^{-3}$) and the density of ZnO ($\rho_{ZnO} = 5.61\ g\ cm^{-3}$), the number of ZnO tetrapods $n_{tetrapods}$ per volume and the surface of tetrapods per volume $VSA_{template}$ can be calculated:

$$V_{tetrapod} = 4 * \frac{\pi * \left(\frac{d}{2}\right)^2 * l}{10^{12}} \qquad (S4)$$

$$n_{tetrapods} = \frac{\rho_{template}}{V_{tetrapod} * \rho_{ZnO}} \qquad (S5)$$

$$VSA_{template} = \frac{\rho_{template}}{V_{tetrapod} * \rho_{ZnO}} \qquad (S6)$$

As the microstructure of aeroglass is composed of hollow microtubes, the $VSA_{AG}$ can be estimated to be twice as the VSA of the template.

$$VSA_{AG} = 2 * VSA_{template} \tag{S7}$$

For microtubes with a rod diameter of 2 μm this results in:

$$VSA_{AG}(2\mu m) = 0.223 \frac{m^2}{cm^3} \tag{S8}$$

The gravimetric surface area GSA is calculated using the AG density $\rho_{AG}$ = 3 mg cm$^{-3}$:

$$GSA_{AG} = \frac{VSA_{AG}}{\rho_{AG}} \tag{S9}$$

For microtubes with a rod diameter of 2 μm this results in:

$$GSA_{AG}(2\mu m) = 74.153 \frac{m^2}{g} \tag{S10}$$

**Supplementary Note 6 – Calculation of surface area 2D vs. 3D**

Assuming an ideal flat 2D surface with a radius of 6 mm, the surface area $A_{2D}$ is calculated by:

$$A_{2D} = \pi \, 0.6 \, cm^2 = 1.13 \, cm^2$$
$$\tag{S11}$$

Considering a cylindrical aeromaterial sample with radius 6 mm and a height of 4 mm, the surface area $A_{3D}$ can be calculated using equations S3 and S5, assuming a tetrapod diameter of 2 μm and a flat surface of the SiO$_2$ microtubes:

$$A_{3D} = A_{tetrapod} * n_{tetrapods} = 503.19 \, cm^2 \tag{S12}$$

**Supplementary Videos**

**Video 1**

Concept of light induced volumetric gas activations in AG/rGO framework structures demonstrated by a gas micro pump. The gas activation (heating) results in a volume expansion, creating a directional flow through the check valves, as visualized by a droplet of dyed water which is pushed through a connected glass tube. The AG/rGO transducer material is illuminated with a frequency of 0.5 Hz as well as 1 Hz and with an irradiance of 1.24 W cm$^{-2}$.

**Video 2**

Demonstration of AG/rGO framework structure as transducer material for applications in light triggered pneumatic (untethered) soft robotics.

**Video 3**

Demonstration of AG/rGO framework structure as transducer material for the application as a light powered membrane pump for fluids.